\documentclass[11pt,aps,prl,amsmath,amssymb,amsfonts,nofootinbib,long]{revtex4}
\usepackage{epsfig,latexsym,bm}

\newcommand\GeV{\mbox{GeV}}

\newcommand\kpc{\mbox{kpc}}
\newcommand\Mpc{\mbox{Mpc}}

\newcommand\G{\mbox{G}}
\newcommand\nG{\mbox{nG}}

\newcommand\B{\mathbf{B}}
\newcommand\E{\mathbf{E}}
\newcommand\x{\mathbf{x}}

\newcommand\mPl{m_{\rm Pl}}
\newcommand\dd{\partial}
\newcommand\rot{\nabla \times}

\begin{document}

\title{Maxwell-Kosteleck\'{y} Electromagnetism and Cosmic Magnetization}

\author{L. Campanelli$^{1,2}$}
\email{leonardo.campanelli@ba.infn.it}
\author{P. Cea$^{1,2}$}
\email{paolo.cea@ba.infn.it}

\affiliation{$^1$Dipartimento di Fisica, Universit\`{a} di Bari, I-70126 Bari, Italy}
\affiliation{$^2$INFN - Sezione di Bari, I-70126 Bari, Italy}

\date{April, 2009}

%************************************   Abstract   *******************************************%

\begin{abstract}
{\bf Abstract}. The Lorentz violating term in the photon sector of
Standard Model Extension, $\mathcal{L}_K = -\mbox{$\frac14$}
(k_F)_{\alpha \beta \mu \nu} F^{\alpha \beta} F^{\mu \nu}$ (here
referred to as the Kosteleck\'{y} term), breaks conformal invariance
of electromagnetism and enables a superadiabatic amplification of
magnetic vacuum fluctuations during inflation. For a wide range of
values of parameters defining Lorentz symmetry violation and
inflation, the present-day magnetic field can have an intensity of
order of nanogauss on megaparsec scales and then could explain the
large-scale magnetization of the universe.
\end{abstract}

%*********************************************************************************************%

%98.62.En -> Electric and magnetic fields
%11.30.Cp -> Lorentz and Poincar\'{e} invariance
%98.80.-k -> Cosmology
\maketitle

%*********************************************************************************************%

% Introduction

\vspace{1.2cm}

The methodical investigation of possible effects of violations of
Lorentz symmetry has became stronger in recent
years~\cite{Kostelecky} (for recent papers concerning Lorentz
violation see, e.g., Refs.~\cite{LVdata,LV}). The interest in
Lorentz violation (LV) resides primarily in the fact that theories
that attempt to unify Gravity to the other fundamental interactions,
such as String Theory or Quantum Gravity, incorporate it in a
natural way. Terrestrial experiments and astrophysical observations,
however, have established that Lorentz violation effects (if ever
they exist) have to be very tiny. Nevertheless, due to the increased
sensitivity of experiments, the detection of LV signals could be a
real possibility in a not-so-far future.

If on the one hand Lorentz symmetry violation is being searched for,
though no compelling evidence for it exists, on the other hand
astrophysical observations undoubtedly indicate that all
large-scale, gravitationally bound systems (i.e. galaxies and
clusters of galaxies) are pervaded by microgauss magnetic fields
whose origin is still not well understood (for reviews on cosmic
magnetic fields see Ref.~\cite{Widrow}). The fact that large-scale
magnetic fields exist everywhere in the universe and possess
approximatively the same intensity seem to indicate that they have a
common and primordial origin. If one takes into account that the
collapse of primordial large-scale structures enhances the intensity
of any preexisting magnetic field of about a factor
$10^3$~\cite{Widrow}, a primeval field with comoving intensity of
order of nanogauss and correlated on megaparsec scales could explain
{\it the magnetization of the universe}.

% Initial Electromagnetic Spectrum from inflation in Maxwell electromagnetism

It is worth noting that, due to the large correlation of the
presently observed fields (ranging from $\sim 10 \kpc$ for
magnetic fields in galaxies to $\sim 1\Mpc$ for those found in
clusters), it is quite natural to suppose that they have been
generated during an inflationary epoch of the universe. Indeed,
during inflation all fields are quantum mechanically exited.
Because the wavelength $\lambda$ associated to a given fluctuation
grows faster than the horizon, there will be a time, say $t_1$,
when this mode crosses outside the horizon itself. After that,
this fluctuation cannot collapse back into the vacuum being not
causally self-correlated, and then ``survives'' as a classical
real object~\cite{Dimopoulos}. The energy associated to a given
fluctuation is subjected to the uncertainty relation, $\Delta E
\Delta t \gtrsim 1$. Therefore, the energy density in the volume
$\Delta V$, $\mathcal{E} = \Delta E/\Delta V$, is approximatively
given by $\mathcal{E} \sim H^4$, where $H$ is the Hubble
parameter. Here, we used the fact that at the horizon crossing
$\Delta t \sim H^{-1}$ and $\Delta V \sim
H^{-3}$~\cite{Dimopoulos}. Taking into account the expression for
the electromagnetic energy in standard Maxwell electromagnetism,
one arrives to the result that the spectrum of magnetic
fluctuations at the time of horizon crossing is given by $B_1 \sim
H^2 \sim M^4/\mPl^2$~\cite{Turner,Dimopoulos,nonlinear}, where in
the last equality we used the Friedmann equation
\begin{equation}
\label{Friedmann} H^2 = (8\pi/3) M^4/\mPl^2.
\end{equation}
Here, $M^4$ is the (constant) total energy density during (de
Sitter) inflation and $\mPl \sim 10^{19} \GeV$ is the Planck mass.

Due to conformal invariance of Maxwell electromagnetism one finds,
however, that the present magnitude of the inflation-produced field
at the scale of, say, $10 \kpc$ is vanishingly small, $B_0 \sim
10^{-52} \G$~\cite{Turner,nonlinear}. Strictly speaking, this is
true only if the background metric is spatially-flat~\cite{Barrow},
which indeed is the case discussed in this paper.

Since the seminal work of Turner and Widrow~\cite{Turner}, a
plethora of mechanisms have been proposed for generating cosmic
magnetic fields in the early universe~\cite{Generation}. Most of
them repose on the breaking of conformal invariance attained by
adding non-conformal invariant terms in the standard electromagnetic
Lagrangian, e.g., non-minimal couplings between photon and gravity,
interaction terms between photon and inflaton, dilaton, or axion,
and so on.

Long time ago, it was noted by Kosteleck\'{y}, Potting and
Samuel~\cite{Kostelecky-p} that the breaking of conformal invariance
is a natural consequence of LV. Indeed, they argued that the
appearance of an effective photon mass owing to spontaneous breaking
of Lorentz invariance could enable the generation of large-scale
magnetic fields within inflationary scenarios. In a subsequent
paper, that idea was developed by Bertolami and
Mota~\cite{Bertolami}. The connection between Lorentz symmetry
violation and cosmic magnetic fields has been studied only in a few
papers, namely within the framework of noncommutative
spacetimes~\cite{Mazumdar,Bamba}, quantum theory with noncommutative
fields~\cite{Gamboa} and, recently, by considering the introduction
of a Lorentz-violating Chern-Simons term in the standard
electromagnetic Lagrangian~\cite{Campanelli}.

%Maxwell-Kosteleck\'{y} action

In this paper, we study the generation of magnetic fields within
the so-called Standard Model Extension (SME)~\cite{Colladay},
which is an effective field theory that includes all admissible
Lorentz-violating terms in the $SU(3) \times SU(2) \times U(1)$
standard model of particle physics. In the photon sector, and in
curved spacetimes, the SME action, here referred to as the
Maxwell-Kosteleck\'{y} action, reads~\cite{Kostelecky1}:
\begin{equation}
\label{Action} S_{\rm MK} = \int \!\! d^4x \, e \! \left[ -
\mbox{$\frac14$} \, F_{\mu \nu} F^{\mu \nu} - \mbox{$\frac14$} \,
(k_F)_{\alpha \beta \mu \nu} F^{\alpha \beta} F^{\mu \nu} \right]
\!,
\end{equation}
where $F_{\mu \nu} = \partial_{\mu} A_{\nu} - \partial_{\nu}
A_{\mu}$ is the electromagnetic field strength tensor and $e$ the
determinant of the vierbein. The presence of the external tensor
$(k_F)_{\alpha \beta \mu \nu}$ breaks, explicitly, (particle)
Lorentz invariance~\cite{Kostelecky1} (and in general conformal
invariance) and parameterizes then Lorentz violation. Note that
$(k_F)_{\alpha \beta \mu \nu}$ is antisymmetric on the first two and
on the last two indices while it is symmetric under interchange of
the first and last pair of indices.

The external tensor $(k_F)_{\alpha \beta \mu \nu}$ is fixed in a
given system of coordinates. Going in different system of
coordinates will, generally, induces a change of the form of
$(k_F)_{\alpha \beta \mu \nu}$. We stress that, in this paper, we
assume that the form of $(k_F)_{\alpha \beta \mu \nu}$ refers to a
system of coordinates at rest with respect to the Cosmic Microwave
Background (CMB), the so-called ``CMB frame''.

Since taking $(k_F)_{\alpha \beta \mu \nu}$ as a fixed tensor
corresponds to have an explicit violation of Lorentz symmetry, this
could introduce in the theory an instability associated to
non-positivity of the energy. This difficulty could be overcome, in
principle, by considering models in which LV is spontaneously
broken, such as those which naturally emerge in string theory. In
this case, however, action~(\ref{Action}) loses its character of
generality, owing to the fact that the tensor $(k_F)_{\alpha \beta
\mu \nu}$ is now regarded as a vacuum expectation value of some
tensor field with its own dynamics. For this reason, we will take
$(k_F)_{\alpha \beta \mu \nu}$ to be a fixed tensor, that is we will
assume that the effects of spontaneous breaking can be approximated
by terms in the action with explicit symmetry breaking. This
assumption is not completely satisfactory from a pure theoretical
viewpoint but, nevertheless, we believe that the ``simplified''
model with explicit symmetry breaking catches the main
characteristics of Lorentz violation in the photon sector.

It should be noted that the photon part of the action for the
Standard Model Extension also contains, in principle, two more
CPT-odd terms: a Chern-Simons term,
\begin{equation}
\label{CSterm} S_{\rm CS} = \int \! d^4x \mbox{$\frac12$} \, e \,
(k_{AF})^\kappa \epsilon_{\kappa \lambda \mu \nu} A^\lambda F^{\mu
\nu},
\end{equation}
and a term linear in the electromagnetic field,
\begin{equation}
\label{Linearterm} S_{\rm A} = -\int \! d^4x \, e \, (k_A)_\kappa
A^\kappa,
\end{equation}
where $\epsilon_{\mu \nu \rho \sigma}$ is the Levi-Civita tensor
density, while $(k_{AF})_\kappa$ and $(k_A)_\kappa$ are coefficients
for Lorentz violation. The first term was analyzed in
Ref.~\cite{Campanelli} while the second one, as it is easy to show,
is not able to break conformal invariance in spatially-flat models
of universe, so there is no production of astrophysically
interesting magnetic fields. For these reasons, we neglect those two
CPT-odd terms in the following analysis.

The equations of motion follow from action~(\ref{Action}):
\begin{equation}
\label{EqMotion} D^{\mu} F_{\mu \nu} + D^{\mu}
[(k_F)_{\mu\nu\alpha\beta} F^{\alpha \beta}] = 0,
\end{equation}
while the Bianchi identities are:
\begin{equation}
\label{Bianchi} D^{\mu} \widetilde{F}_{\mu\nu} = 0,
\end{equation}
where $D_\mu$ is the spacetime covariant derivative and
\begin{equation}
\label{dualF} \widetilde{F}_{\mu \nu} = \frac{1}{2e} \,
\epsilon_{\mu\nu\alpha\beta} F^{\alpha\beta}
\end{equation}
is the dual electromagnetic field strength tensor.

We assume that the universe is described by a Robertson-Walker
metric
\begin{equation}
\label{RWmetric} ds^2 = a^2(d\eta^2 - d \x^2),
\end{equation}
where $a(\eta)$ is the expansion parameter, $\eta$ the conformal
time (related to the cosmic time $t$ through $d\eta = dt/a$), and
$H$ the Hubble parameter. Introducing, in the usual way, the
electric and magnetic fields as $F_{0i} = -a^2 E_i$ and $F_{ij} =
\epsilon_{ijk} a^2 B_k$ (Latin indices run from $1$ to $3$, while
Greek ones from $0$ to $3$), the equations of motion read:
\begin{eqnarray}
\label{motion} && \partial_\eta (a^2 E_i) - \epsilon_{ijk}
\partial_j (a^2 B_k) + \nonumber \\
&& \partial_\eta [2 (k_F)_{i00j} a^{-2} E_j -
(k_F)_{0ijk} \epsilon_{jkl} a^{-2} B_l] + \nonumber \\
&& \partial_j [2 (k_F)_{ijk0} a^{-2} E_k - (k_F)_{ijkl}
\epsilon_{klm} a^{-2} B_m] = 0, \\ \nonumber \\
&& \partial_i (a^2 E_i) +
\partial_i[ 2 (k_F)_{i00j} a^{-2} E_j - (k_F)_{0ijk} \epsilon_{jkl}
a^{-2} B_l] = 0,
\end{eqnarray}
while the Bianchi identities become:
\begin{equation}
\label{Bianchi2}
\partial_\eta(a^2 {\textbf B}) + \nabla \times (a^2 {\textbf E}) =
0, \;\;\; \nabla \cdot {\textbf B} = 0.
\end{equation}
Moreover, following the standard procedure, we find the
electromagnetic energy-momentum tensor:
\begin{equation}
\label{T} T_{\mu\nu} = \frac14 \, g_{\mu\nu}
F_{\alpha\beta}F^{\alpha\beta} - F_\mu^{\;\;\alpha}F_{\nu\alpha} +
\frac14 \, g_{\mu\nu} (k_F)_{\alpha\beta\gamma\delta}
F^{\alpha\beta} F^{\gamma\delta} - (k_F)_{\mu\alpha\beta\gamma}
F_\nu^{\;\;\alpha} F^{\beta\gamma}.
\end{equation}
The electromagnetic energy density is then given by
\begin{equation}
\label{T00} \mathcal{E} = T^0_{\;\;0} = \frac12 ({\textbf E}^2 +
{\textbf B}^2) + (k_F)_{i00j} a^{-4} E_i E_j + \frac14 (k_F)_{ijkl}
a^{-4} \epsilon_{ijm} \epsilon_{kln} B_m B_n.
\end{equation}
In any sensible Lorentz-violating theory, we require positivity of
the energy. In the Maxwell-Kosteleck\'{y} theory, the energy density
is not-positive defined in general. If in a particular system of
coordinates the energy in not-positive defined, this simply means
that this effective theory becomes meaningless and one needs to
consider the full theory with spontaneous Lorentz symmetry breaking
in order to get physically acceptable results.

% Evolution of inflation-produced fields during inflation and reheating

We are interested in the evolution of electromagnetic fields well
outside the horizon, that is to modes whose physical wavelength is
much greater than the Hubble radius $H^{-1}$, $\lambda_{\rm phys}
\gg H^{-1}$, where $\lambda_{\rm phys} = a \lambda$ and $\lambda$
is the comoving wavelength. Since $a \eta \sim H^{-1}$,
introducing the comoving wavenumber $k = 2\pi/\lambda$, the above
condition reads $|k\eta| \ll 1$. Observing that the first Bianchi
identity gives $B \sim k\eta E$, where $B$ and $E$ stand for the
average magnitude of the magnetic and electric field intensities,
and assuming that all (non-null) components of $(k_F)_{\alpha
\beta \mu \nu}$ have approximatively the same magnitude,
it easy to see that, at large scales, Eq.~(\ref{motion}) reduces
to
%
%we can neglect, on large scales, the second terms with respect to
%the first ones in each line of Eq.~(\ref{motion}). Also, the first
%term in third line is negligible with respect to the first one in
%the second line. Therefore, at large scales, Eq.~(\ref{motion})
%reduces to
%
\begin{equation}
\label{motionLS}
\partial_\eta (a^2 E_i) +
\partial_\eta [2 (k_F)_{i00j} a^{-2} E_j] = 0.
\end{equation}
Assuming that $||(k_F)_{i00j}|| \gg a^4$ we then have $(k_F)_{i00j}
a^{-2} E_j = c_i$, where the $c_i$'s are constants of integration.
For the sake of simplicity, we assume that $(k_F)_{i00j}$ is a
constant isotropic tensor, so that we can write
\begin{equation}
\label{k} (k_F)_{i00j} = k_F \delta_{ij},
\end{equation}
where $k_F$ is a constant which gives the magnitude of Lorentz
violation effect, and $\delta_{ij}$ is the Kronecker delta. In this
case, the average intensity of the electric field scales as $E
\propto a^2$. Observing that $\eta \propto a^{-1}$ during de Sitter
inflation and taking into account the first Bianchi identity, we get
that the average magnetic intensity grows ``superadiabatically'', $B
\propto a$.
%[We have assumed that in the CMB frame $(k_F)_{i00j}$ is different
%from zero. If $(k_F)_{i00j}$ is zero or negligibly small, the
%inflation-produced field is, in general, small-scale correlated
%and then not astrophysically interesting.]

% Initial Electromagnetic Spectrum from inflation in Maxwell-Kosteleck\'{y} electromagnetism

Before proceeding further, we estimate the spectrum of magnetic
vacuum fluctuation generated during de Sitter inflation in
Maxwell-Kosteleck\'{y} electromagnetism. If, for the sake of
simplicity, we assume that the nonzero components of $(k_F)_{\alpha
\beta \mu \nu}$ are just given by Eq.~(\ref{k}), the electromagnetic
energy density turns to be
\begin{equation}
\label{Energy} \mathcal{E} = \frac12 ({\textbf E}^2 + {\textbf B}^2)
+ k_F a^{-4} {\textbf E}^2.
\end{equation}
As a consequence, at the time of crossing, where $|k\eta| \sim 1$,
and in the limit $k_F \gg a^4$, we get $\mathcal{E} \sim k_F
a_1^{-4} B_1^2$, where $a_1 = a(t_1)$ and we used the first Bianchi
identity. Remembering that $\mathcal{E} \sim H^4$, we obtain the
spectrum of magnetic fluctuations at the time of horizon crossing:
\begin{equation}
\label{B1} B_1 \sim k_F^{-1/2} a_1^2 H^2.
\end{equation}
It is worth noting that, in order to have positivity of the energy,
we are forced to assume $k_F > 0$.
%[If we consider the full expression for the electromagnetic energy
%density, and assume that it is positive defined, we find that at
%the crossing and in the limit $||(k_F)_{\alpha\beta\mu\nu}|| \gg
%a$ the expression of $\mathcal{E}$ is still given by $\mathcal{E}
%\sim k_F a_1^{-4}B_1^2$, but, in this case, $k_F$ stands for $k_F
%\sim \max ||(k_F)_{\alpha\beta\mu\nu}||$.]

% Evolution of inflation-produced fields during reheating

After inflation, the universe enters in the so-called reheating
phase, during which the energy of the inflaton is converted into
ordinary matter. The reheating phase ends at the temperature $T_{\rm
RH}$ which is less than $M$ and constrained as~\cite{Riotto}
\begin{equation}
\label{TRH} T_{\rm RH} \lesssim 10^8 \GeV.
\end{equation}
Moreover, CMB analysis requires $M \lesssim
10^{-2}\mPl$~\cite{Turner}, otherwise it would be too much of a
gravitational waves relic abundance, and also one must impose that
$T_{\rm RH} \gtrsim 1\GeV$, so that the predictions of Big Bang
Nucleosynthesis (BBN) are not spoiled~\cite{Turner}.

It is worth noting that the condition $||(k_F)_{i00j}|| \gg a^4$
or, equivalently, $k_F \gg a^4$ is certainly fulfil during
inflation and reheating if $k_F \gg a_{\rm RH}^4$, where $a_{\rm
RH} = a(T_{\rm RH})$. Since $a_{\rm RH} \sim T_0/T_{\rm RH}$,
where $T_0 \sim 10^{-13} \GeV$ is the actual
temperature~\cite{Kolb}, we have $10^{-84} \lesssim a_{\rm RH}^4
\lesssim$ $10^{-52}$ for $1\GeV \lesssim T_{\rm RH} \lesssim 10^8
\GeV$.

The most stringent upper bounds on $||(k_F)_{\alpha \beta \mu
\nu}||$ come from the analysis of CMB polarization and polarized
light of radiogalaxies and gamma-ray bursts, and are respectively:
$10^{-30}$~\cite{Kostelecky-m1}, $10^{-32}$~\cite{Kostelecky-m2},
and $10^{-37}$~\cite{Kostelecky-m3},
\footnote{It should be noted that, though the CMB bound is less
stringent than the radiogalaxies and gamma-ray bursts bounds, it
covers the whole portion of coefficient space for Lorentz violation.
The point-source nature of radiogalaxies and gamma-ray bursts,
instead, allows us to put constraints only on limited portions of
coefficient space~\cite{Kostelecky-m1}.}
although, in deriving these constraints, some additional assumptions
have to be taken. A direct constraint on the quantity $k_F$ [defined
through Eq.~(\ref{k})] can be obtained if one takes into account
that the maximal attained experimental sensitivity on the
coefficient $\tilde{\kappa}_{\rm tr} \equiv 2k_F$ introduced in
Ref.~\cite{LVdata} is about $10^{-11}$ and that there is no
compelling evidence for nonzero values of $\tilde{\kappa}_{\rm tr}$
(see the 2009 version of the Data Tables for Lorentz and CPT
Violation~\cite{LVdata} for more details).
\\
In any case, for a wide range of allowed values of $k_F$, we have a
superadiabatic amplification of magnetic vacuum fluctuations during
inflation and reheating. In this latter phase, in particular, taking
into account that $\eta \propto a^{1/2}$, we have $B \propto
a^{5/2}$.

% Plasma Effects

After reheating, the universe enters the radiation dominated era.
In this era, as well as in the subsequent matter era, the effects
of the conducting primordial plasma are important when studying
the evolution of a magnetic field. They are taken into account by
adding to the electromagnetic Lagrangian the source term $j^{\mu}
\! A_{\mu}$~\cite{Turner}. Here, the external current $j^{\mu}$,
expressed in terms of the electric field, has the form $j^{\mu} =
(0, \sigma_c {\textbf E})$, where $\sigma_c$ is the conductivity.
Plasma effects introduce, in the right-hand-side of
Eq.~(\ref{motion}), the extra term $-a\sigma_c (a^2 E_i)$. In this
case, it easy to see that modes well outside the horizon (assuming
that $k_F \gg a^4$) evolve as
\begin{equation}
\label{E} E \propto a^2 \exp \!\left(-\!\int \!d\eta \,
a^5\sigma_c/2k_F\right) \! .
\end{equation}
Approximating $\int \!d\eta \, a^5\sigma_c$ with $\eta \,
a^5\sigma_c$ and using $a\eta \sim H^{-1}$, we get from
Eq.~(\ref{E}):
\begin{equation}
\label{Eapprox} E \propto a^2 \exp(-a^4\sigma_c/2Hk_F).
\end{equation}
In the radiation era $H \sim T^2/\mPl$~\cite{Kolb} and, for
temperature much greater than the electron mass, the conductivity is
approximatively given by $\sigma_c \sim T/\alpha$~\cite{Turner},
where $\alpha$ is the fine structure constant and $T$ the
temperature. Then, from Eq.~(\ref{Eapprox}) we obtain:
\begin{equation}
\label{Esol} E \propto a^2 \exp[-(T_*/T)^5)],
\end{equation}
where $T_*/10\GeV \sim (10^{-37}\!/k_F)^{1/5}$. This means that for
$T \gtrsim T_*$ we have $E \propto a^2$ (which in turn gives $B
\propto a^3$ since $\eta \propto a$ in radiation era), while for $T
\lesssim T_*$ the electric field is dissipated, so the magnetic
field evolves adiabatically, $B \propto a^{-2}$. [We have assumed
that $k_F \gg a^4$ from the end of reheating until $T_*$. This
assumption is satisfied if $k_F \gtrsim a_*^4$, where $a_* = a(T_*)$
or, taking into account the above expression for $T_*$, if $k_F
\gtrsim 10^{-95}$. This is certainly our case since we have already
assumed that $k_F \gtrsim 10^{-52}$ if, for example, $T_{\rm RH}
\simeq 1\GeV$ or $k_F \gtrsim 10^{-84}$ if $T_{\rm RH} \simeq 10^8
\GeV$.]

% Actual magnetic field

Finally, evolving along the lines discussed above the
inflation-produced magnetic field from the time of
horizon crossing until today, %($a=a_0$),
we get
\begin{equation}
\label{B0} \frac{B_0}{\nG} \sim \left(\frac{M}{10^{14} \GeV}
\right)^{\!\!4} \left(\frac{T_{\rm RH}}{10 \GeV}
\right)^{\!\!5(n-1)/2} \!\! \left(\frac{k_F}{10^{-37}}
\right)^{\!\!n/2} \lambda_{\rm Mpc}^{-1} \, ,
\end{equation}
%
%\begin{equation}
%\label{B0} B_0 \sim \left(\frac{M}{10^{14} \GeV} \right)^{\!\!4}
%\! \left(\frac{T_{\rm RH}}{10 \GeV} \right)^{\!\!\frac{5(n-1)}{2}}
%\! \left(\frac{k_F}{10^{-37}} \right)^{\!\!\frac n2} \lambda_{\rm
%Mpc}^{-1} \, \nG,
%\end{equation}
%
where $n$ takes the values $\pm 1$ according to $T_* \lesssim
T_{\rm RH}$ or $T_* \gtrsim T_{\rm RH}$. The latter condition
means, indeed, that the magnetic field evolves adiabatically from
the end of reheating until today and is equivalent to have $T_{\rm
RH}/10\GeV \lesssim (10^{-37}\!/k_F)^{1/5}$.
%
%\begin{eqnarray}
%\label{B0} B_0 \!\!& \sim &\!\! B_1 \left( \frac{a_{\rm end}}{a_1}
%\right) \! \left(\frac{a_{\rm RH}}{a_{\rm end}}\right)^{\!\!5/2}
%\!\! \left(\frac{a_*}{a_{\rm RH}}\right)^{\!\!\frac{5n+1}{2}} \!
%\left(\frac{a_0}{a_*} \right)^{\!\!-2} \nonumber \\
%\!\!& \sim &\!\! \left(\frac{M}{10^{14} \GeV} \right)^{\!\!4} \!
%\left(\frac{T_{\rm RH}}{10 \GeV} \right)^{\!\!\frac{5(n-1)}{2}} \!
%\left(\frac{k_F}{10^{-37}}
%\right)^{\!\!\frac n2} \lambda_{\rm Mpc}^{-1} \, \nG, \nonumber \\
%\end{eqnarray}
%
%where $a_{\rm end}$ is the value of $a$ at the end of inflation,
%$\lambda_{\rm Mpc} = \lambda/\Mpc$, and we used $a_{\rm end}/a_1 =
%e^{N(\lambda)}$, with $N(\lambda) \simeq 38 + (2/3)\ln
%(M/10^{14}\GeV) + (1/3) \ln (T_{\rm RH}/10\GeV) + \ln \lambda_{\rm
%Mpc}$ the number of $e$-folds elapsing from the crossing of a
%comoving length $\lambda$ outside the horizon to the end of
%inflation~\cite{Turner}. In obtaining Eq.~(\ref{B0}), we also used
%the fact that during reheating $\rho_{\rm tot} \propto a^{-3}$,
%where $\rho_{\rm tot}$ is the total energy density of the
%universe, and that $\rho_{\rm tot}(a_{\rm end}) \sim M^4$ and
%$\rho_{\rm tot}(a_{\rm RH}) \sim T_{\rm RH}^4$. The exponent $n$
%takes the values $\pm 1$ according to $T_* \lesssim T_{\rm RH}$ or
%$T_* \gtrsim T_{\rm RH}$. The latter condition means, indeed, that
%the magnetic field evolves adiabatically from the end of reheating
%until today and is equivalent to have $T_{\rm RH}/10\GeV \lesssim
%(10^{-37}\!/k_F)^{1/5}$.

It is clear from Eq.~(\ref{B0}) and Fig.~1 that, for a wide range of
values of parameters defining inflation ($M$ and $T_{\rm RH}$) and
Lorentz symmetry violation ($10^{-52} \lesssim k_F \lesssim
10^{-11}$), the present-day magnetic field can be as strong as $B_0
\sim 1.0 \, \nG$ on megaparsec scales.

%************************************   Figure 1   *******************************************%

\begin{figure}[t]
\begin{center}
\includegraphics[clip,width=0.65\textwidth]{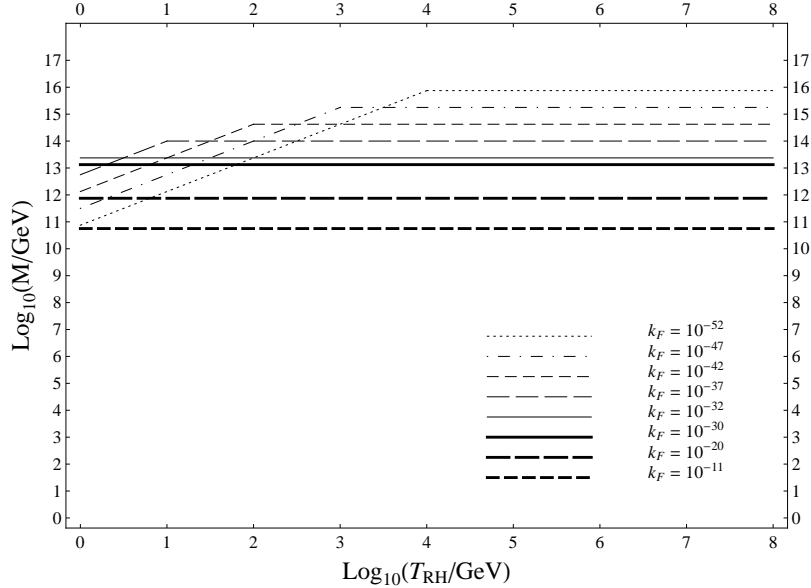}
\caption{The inflation-produced magnetic field has an actual
intensity of order of nanogauss on megaparsec scales if the values
of parameter defining inflation, i.e. energy scale of inflation
$M$ and the reheat temperature $T_{\rm RH}$, stay on the curves.
$k_F \sim ||(k_F)_{\alpha \beta \mu \nu}||$ estimates the
magnitude of the (constant) external tensor which parameterizes
Lorentz violation [see Eq.~(\ref{Action})].}
\end{center}
\end{figure}

%*********************************************************************************************%

% Conclusions

In conclusion, we have shown that the presence of large-scale
magnetic fields in the present universe is a general prediction of
Standard Model Extension. The presence of the ``Kosteleck\'{y}'',
Lorentz violating term $\mathcal{L}_K = -\mbox{$\frac14$}
(k_F)_{\alpha \beta \mu \nu} F^{\alpha \beta} F^{\mu \nu}$ in the
photon sector of the theory allows electromagnetic vacuum
fluctuations to be superadiabatically amplified during inflation.
The resulting magnetic field has, today, a magnitude of order of
nanogauss on megaparsec scales for a wide range of parameter space
of inflation and Lorentz violation and then could explain why our
universe is magnetized.

%********************************   Acknowledgments   ****************************************%

\begin{acknowledgments}

\end{acknowledgments}

%**********************************   Bibliography   *****************************************%


\begin{thebibliography}{99}

\bibitem{Kostelecky}     V.~A.~Kosteleck\'{y},
                         %``Perspectives on Lorentz and CPT Violation,''
                         arXiv:0802.0581 [gr-qc].

\bibitem{LVdata}         V.~A.~Kosteleck\'{y} and N.~Russell,
                         %``Data Tables for Lorentz and CPT Violation,''
                         arXiv:0801.0287 [hep-ph].

\bibitem{LV}             E.~Komatsu {\it et al.} (WMAP Collaboration),
                         %``Five-Year Wilkinson Microwave Anisotropy Probe (WMAP\altaffilmark 1 )
                         %Observations:Cosmological Interpretation,''
                         arXiv:0803.0547 [astro-ph];
                         R.~K.~Obousy and G.~Cleaver,
                         %``Radius Destabilization in Five Dimensional Orbifolds
                         %from Lorentz Violating Fields,''
                         arXiv:0805.0019 [gr-qc].

\bibitem{Widrow}         L.~M.~Widrow,
                         %``Origin of Galactic and Extragalactic Magnetic Fields,''
                         Rev.\ Mod.\ Phys.\ {\bf 74}, 775 (2002);
                         M.~Giovannini,
                         %``The magnetized universe,''
                         Int.\ J.\ Mod.\ Phys.\ D {\bf 13}, 391 (2004);
                         D.~Grasso and H.~R.~Rubinstein,
                         %``Magnetic fields in the early universe,''
                         Phys.\ Rept.\ {\bf 348}, 163 (2001).

\bibitem{Dimopoulos}     K.~Dimopoulos,
                         %``Galactic magnetic fields as a consequence of inflation,''
                         astro-ph/0105488 (unpublished).

\bibitem{nonlinear}      L.~Campanelli, P.~Cea, G.~L.~Fogli and L.~Tedesco,
                         %``Inflation-Produced Magnetic Fields in Nonlinear Electrodynamics,''
                         Phys.\ Rev.\ D {\bf 77}, 043001 (2008).

\bibitem{Turner}         M.~S.~Turner and L.~M.~Widrow,
                         %``INFLATION PRODUCED, LARGE SCALE MAGNETIC FIELDS,''
                         Phys.\ Rev.\ D {\bf 37}, 2743 (1988).

\bibitem{Barrow}         J.~D.~Barrow and C.~G.~Tsagas,
                         %``Slow decay of magnetic fields in open Friedmann universes,''
                         arXiv:0803.0660 [astro-ph].

\bibitem{Generation}     An incomplete list of mechanisms for generating cosmic
                         magnetic fields is:
                         B.~Ratra,
                         %``Cosmological 'seed' magnetic field from inflation,''
                         Astrophys.\ J.\ {\bf 391}, L1 (1992);
                         W.~D.~Garretson, G.~B.~Field and S.~M.~Carroll,
                         %``Primordial magnetic fields from pseudoGoldstone bosons,''
                         Phys.\ Rev.\ D {\bf 46}, 5346 (1992);
                         A.~Dolgov,
                         %``Breaking Of Conformal Invariance And Electromagnetic Field Generation In
                         %The Universe,''
                         {\it ibid.} {\bf 48}, 2499 (1993);
                         F.~D.~Mazzitelli and F.~M.~Spedalieri,
                         %``Scalar electrodynamics and primordial magnetic fields,''
                         {\it ibid.} {\bf 52}, 6694 (1995);
                         D.~Lemoine and M.~Lemoine,
                         %``Primordial magnetic fields in string cosmology,''
                         {\it ibid.} {\bf 52}, 1955 (1995);
                         M.~Gasperini, M.~Giovannini and G.~Veneziano,
                         %``Primordial magnetic fields from string cosmology,''
                         Phys.\ Rev.\ Lett.\ {\bf 75}, 3796 (1995);
                         A.~C.~Davis and K.~Dimopoulos,
                         %``Primordial Magnetic Fields In False Vacuum Inflation,''
                         Phys.\ Rev.\ D {\bf 55}, 7398 (1997);
                         J.~M.~Cornwall,
                         %``Speculations on primordial magnetic helicity,''
                         {\it ibid.} {\bf 56}, 6146 (1997);
                         M.~Giovannini and M.~E.~Shaposhnikov,
                         %``Primordial magnetic fields, anomalous isocurvature fluctuations and big
                         %bang nucleosynthesis,''
                         Phys.\ Rev.\ Lett.\ {\bf 80}, 22 (1998);
                         %``Primordial hypermagnetic fields and triangle anomaly,''
                         Phys.\ Rev.\ D {\bf 57}, 2186 (1998);
                         A.~Berera, T.~W.~Kephart and S.~D.~Wick,
                         %``GUT cosmic magnetic fields in a warm inflationary universe,''
                         {\it ibid.} {\bf 59}, 043510 (1999);
                         M.~Giovannini,
                         %``Primordial hypermagnetic knots,''
                         {\it ibid.} {\bf 61}, 063004 (2000);
                         %``Hypermagnetic knots, Chern-Simons waves and the baryon asymmetry,''
                         {\bf 61}, 063502 (2000);
                         %``Magnetogenesis and the dynamics of internal dimensions,''
                         {\bf 62}, 123505 (2000);
                         %``On the variation of the gauge couplings during inflation,''
                         {\bf 64}, 061301 (2001);
                         G.~B.~Field and S.~M.~Carroll,
                         %``Cosmological magnetic fields from primordial helicity,''
                         {\it ibid.} {\bf 62}, 103008 (2000);
                         M.~Gasperini,
                         %``A new mechanism for the generation of primordial seeds for the cosmic
                         %magnetic fields,''
                         Phys.\ Rev.\ D {\bf 63}, 047301 (2001);
                         T.~Prokopec,
                         %``Cosmological magnetic fields from photon coupling to fermions and  bosons
                         %in inflation,''
                         astro-ph/0106247 (unpublished);
                         T.~Vachaspati,
                         %``Estimate of the primordial magnetic field helicity,''
                         Phys.\ Rev.\ Lett.\ {\bf 87}, 251302 (2001);
                         B.~A.~Bassett {\it et al.}, %G.~Pollifrone, S.~Tsujikawa, and F.~Viniegra,
                         %``Preheating as cosmic magnetic dynamo,''
                         Phys.\ Rev.\ D {\bf 63}, 103515 (2001);
                         K.~Dimopoulos {\it et al.}, %T.~Prokopec, O.~Tornkvist, and A.~C.~Davis,
                         %``Natural magnetogenesis from inflation,''
                         {\it ibid.} {\bf 65}, 063505 (2002);
                         M.~Marklund {\it et al.}, %P.~K.~S.~Dunsby, M.~Servin, G.~Betschart and C.~Tsagas,
                         %``Charged multifluids in general relativity,''
                         Class.\ Quant.\ Grav.\ {\bf 20}, 1823 (2003);
                         G.~Betschart, P.~K.~S.~Dunsby and M.~Marklund,
                         %``Cosmic magnetic fields from velocity perturbations in the early Universe,''
                         {\it ibid.} {\bf 21}, 2115 (2004);
                         C.~Zunckel {\it et al.}, %G.~Betschart, P.~K.~S.~Dunsby and M.~Marklund,
                         %``On inhomogeneous magnetic seed fields and gravitational waves within  the
                         %MHD limit,''
                         Phys.\ Rev.\ D {\bf 73}, 103509 (2006);
                         K.~Bamba and J.~Yokoyama,
                         %``Large-scale magnetic fields from inflation in dilaton electromagnetism,''
                         {\it ibid.} {\bf 69}, 043507 (2004);
                         T.~Prokopec and E.~Puchwein,
                         %``Nearly minimal magnetogenesis,''
                         {\it ibid.} {\bf 70}, 043004 (2004);
                         A.~Ashoorioon and R.~B.~Mann,
                         %``Generation of cosmological seed magnetic fields from inflation with
                         %cutoff,''
                         {\it ibid.} {\bf 71}, 103509 (2005);
                         C.~G.~Tsagas,
                         %``Resonant amplification of magnetic seed fields by gravitational waves in
                         %the early universe,''
                         {\it ibid.} {\bf 72}, 123509 (2005);
                         L.~Campanelli and M.~Giannotti,
                         %``Magnetic helicity generation from the cosmic axion field,''
                         {\it ibid.} {\bf 72}, 123001 (2005);
                         M.~R.~Garousi, M.~Sami and S.~Tsujikawa,
                         %``Generation of electromagnetic fields in string cosmology with a massive
                         %scalar field on the anti D-brane,''
                         Phys.\ Lett.\ B {\bf 606}, 1 (2005);
                         M.~Laine,
                         %``Real-time Chern-Simons term for hypermagnetic fields,''
                         JHEP {\bf 0510}, 056 (2005);
                         M.~M.~Anber and L.~Sorbo,
                         %``N-flationary magnetic fields,''
                         JCAP {\bf 0610}, 018 (2006);
                         K.~Bamba and M.~Sasaki,
                         %``Large-scale magnetic fields in the inflationary universe,''
                         {\it ibid.} {\bf 0702}, 030 (2007);
                         J.~Martin and J.~Yokoyama,
                         %``Generation of Large-Scale Magnetic Fields in Single-Field Inflation,''
                         {\it ibid.} {\bf 0801}, 025 (2008);
                         A.~Akhtari-Zavareh, A.~Hojati and B.~Mirza,
                         %``Generation of large scale magnetic fields by coupling to curvature and
                         %dilaton field,''
                         Prog.\ Theor.\ Phys.\ {\bf 117}, 803 (2007);
                         %[arXiv:0707.3493 [astro-ph]];
                         K.~E.~Kunze,
                         %``Primordial magnetic seed fields from extra dimensions,''
                         Phys.\ Lett.\ B {\bf 623}, 1 (2005);
                         %``Primordial magnetic fields and nonlinear electrodynamics,''
                         Phys.\ Rev.\ D {\bf 77}, 023530 (2008);
                         M.~Giovannini,
                         %``Magnetogenesis, spectator fields and CMB signatures,''
                         Phys.\ Lett.\ B {\bf 659}, 661 (2008);
                         A.~Diaz-Gil {\it et al.}, %J.~Garcia-Bellido, M.~Garcia Perez and A.~Gonzalez-Arroyo,
                         %``Magnetic field production after inflation,''
                         PoS {\bf LAT2005}, 242 (2006) [arXiv:hep-lat/0509094];
                         %``Magnetic field production during preheating at the electroweak scale,''
                         Phys.\ Rev.\ Lett.\ {\bf 100}, 241301 (2008);
                         %``Primordial magnetic fields from preheating at the electroweak scale,''
                         JHEP {\bf 0807}, 043 (2008);
                         K.~Bamba,
                         %``The interrelation between the generation of large-scale electric fields and
                         %that of large-scale magnetic fields during inflation,''
                         JCAP {\bf 0710}, 015 (2007);
                         K.~Bamba and S.~D.~Odintsov,
                         %``Inflation and late-time cosmic acceleration in non-minimal Maxwell-$F(R)$
                         %gravity and the generation of large-scale magnetic fields,''
                         {\it ibid.} {\bf 0804}, 024 (2008);
                         K.~Bamba, N.~Ohta and S.~Tsujikawa,
                         %``Generic estimates for magnetic fields generated during inflation including
                         %Dirac-Born-Infeld theories,''
                         Phys.\ Rev.\ D {\bf 78}, 043524 (2008);
                         K.~Bamba, S.~Nojiri and S.~D.~Odintsov,
                         %``Inflationary cosmology and the late-time accelerated expansion of the
                         %universe in non-minimal Yang-Mills-$F(R)$ gravity and non-minimal
                         %vector-$F(R)$ gravity,''
                         {\it ibid.} {\bf 77}, 123532 (2008);
                         C.~J.~Copi {\it et al.}, %F.~Ferrer, T.~Vachaspati, and A.~Achucarro,
                         %``Helical Magnetic Fields from Sphaleron Decay and Baryogenesis,''
                         Phys.\ Rev.\ Lett.\ {\bf 101}, 171302 (2008);
                         L.~Campanelli, P.~Cea, G.~L.~Fogli and L.~Tedesco,
                         %``Inflation-Produced Magnetic Fields in R^n F^2 and I F^2 models,''
                         Phys.\ Rev.\ D {\bf 77}, 123002 (2008);
                         L.~Campanelli,
                         %Helical Magnetic Fields from Inflation,
                         arXiv:0805.0575 [astro-ph];
                         K.~Bamba, C.~Q.~Geng and S.~H.~Ho,
                         %``Large-scale magnetic fields from inflation due to Chern-Simons-like
                         %effective interaction,''
                         JCAP {\bf 0811}, 013 (2008).

\bibitem{Kostelecky-p}   V.~A.~Kosteleck\'{y}, R.~Potting and S.~Samuel,
                         %``String signatures''
                         in {\it Proceedings of the 1991 Joint International Lepton-Photon
                         Symposium and Europhysics Conference on High Energy Physics},
                         %Geneva, Switzerland,
                         edited by S.~Hegarty, K.~Potter, E.~Quercigh
                         (World Scientific, Singapore, 1992).

\bibitem{Bertolami}      O.~Bertolami and D.~F.~Mota,
                         %``Primordial magnetic fields via spontaneous breaking of Lorentz
                         %invariance,''
                         Phys.\ Lett.\ B {\bf 455}, 96 (1999).

\bibitem{Mazumdar}       A.~Mazumdar and M.~M.~Sheikh-Jabbari,
                         %``Noncommutativity in space and primordial magnetic field,''
                         Phys.\ Rev.\ Lett.\ {\bf 87}, 011301 (2001).

\bibitem{Bamba}          K.~Bamba and J.~Yokoyama,
                         %``Large-scale magnetic fields from dilaton inflation in noncommutative
                         %spacetime,''
                         Phys.\ Rev.\ D {\bf 70}, 083508 (2004).

\bibitem{Gamboa}         J.~Gamboa and J.~Lopez-Sarrion,
                         %``U(1) noncommutative gauge fields and magnetogenesis,''
                         Phys.\ Rev.\ D {\bf 71}, 067702 (2005).

\bibitem{Campanelli}     L.~Campanelli, P.~Cea and G.~L.~Fogli,
                         %``Lorentz Symmetry Violation and Galactic Magnetism,''
                         arXiv:0805.1851 [astro-ph].

\bibitem{Colladay}       D.~Colladay and V.~A.~Kosteleck\'{y},
                         %``CPT violation and the standard model,''
                         Phys.\ Rev.\ D {\bf 55}, 6760 (1997);
                         %``Lorentz-violating extension of the standard model,''
                         {\bf 58}, 116002 (1998).

\bibitem{Kostelecky1}    V.~A.~Kosteleck\'{y},
                         %``Gravity, Lorentz violation, and the standard model,''
                         Phys.\ Rev.\ D {\bf 69}, 105009 (2004);
                         %``Lorentz violation and gravity,''
                         arXiv:hep-ph/0412406.

\bibitem{Riotto}         For reviews on Inflation and BBN see, respectively:
                         D.~H.~Lyth and A.~Riotto,
                         %``Particle physics models of inflation and the cosmological density
                         %perturbation,''
                         Phys.\ Rept.\ {\bf 314}, 1 (1999);
                         S.~Sarkar,
                         %``Big bang nucleosynthesis and physics beyond the standard model,''
                         Rept.\ Prog.\ Phys.\ {\bf 59}, 1493 (1996).

\bibitem{Kolb}           E.~W.~Kolb and M.~S.~Turner,
                         {\it The Early Universe}
                         (Addison-Wesley, Redwood City, CA, 1990).

\bibitem{Kostelecky-m1}  V.~A.~Kosteleck\'{y} and M.~Mewes,
                         %``Lorentz-violating electrodynamics and the cosmic microwave background,''
                         Phys.\ Rev.\ Lett.\ {\bf 99}, 011601  (2007).

\bibitem{Kostelecky-m2}  V.~A.~Kosteleck\'{y} and M.~Mewes,
                         %``Cosmological constraints on Lorentz violation in electrodynamics,''
                         Phys.\ Rev.\ Lett.\ {\bf 87}, 251304 (2001);
                         %``Signals for Lorentz violation in electrodynamics,''
                         Phys.\ Rev.\ D {\bf 66}, 056005 (2002).

\bibitem{Kostelecky-m3}  V.~A.~Kosteleck\'{y} and M.~Mewes,
                         %``Sensitive polarimetric search for relativity violations in gamma-ray
                         %bursts,''
                         Phys.\ Rev.\ Lett.\ {\bf 97}, 140401 (2006).

\end{thebibliography}
\end{document}